\def\final {1}	

\ifdefined\preprint
  \documentclass[preprint,review,12pt]{elsarticle}
\fi
\ifdefined\wordcount
  \documentclass[final,3p,times,twocolumn]{elsarticle}
\fi
\ifdefined\final
  \documentclass[final,3p,times,twocolumn]{elsarticle}
\fi

\usepackage{graphicx,stfloats}
\usepackage{color}
\graphicspath{{figs/}}
\usepackage[version=4]{mhchem}
\usepackage{xspace}
\usepackage{booktabs}

\biboptions{sort&compress}
\journal{Proceedings of the Combustion Institute}

\newcommand{\ie}{\textit{i.e.},\xspace}
\newcommand{\eg}{\textit{e.g.},\xspace}

\newcommand{\tit}{The Role of Molecular Properties on the Dimerization of Aromatic Compounds}

\begin{document}

\begin{frontmatter}

\title{\tit}

\author[fir]{Paolo Elvati}
\author[fir]{Kirk Turrentine}
\author[fir,sec]{Angela Violi}

\cortext[cor1]{Corresponding author:}
\ead{avioli@umich.edu}

\address[fir]{Department of Mechanical Engineering, University of Michigan, Ann Arbor, MI 48109-2125, USA}
\address[sec]{Departments of Chemical Engineering, Biomedical Engineering, Macromolecular Science and Engineering, Biophysics Program, University of Michigan, Ann Arbor, MI, United States}

\begin{abstract}
Recent results have shown the presence and importance of oxygen chemistry during the growth of aromatic compounds, leading to the formation of oxygenated structures that have been identified in various environments. Since the formation of polycyclic aromatic compounds (PAC) bridge the formation of gas-phase species with particle inception, in this work we report a detailed analysis of the effects of molecular characteristics on physical growth of PAC via dimerization. We have included oxygen content, mass, type of bonds (rigid versus rotatable), and shape as main properties of the molecules and studied their effect on the propensity of these structures to form homo-molecular and hetero-molecular dimers. Using enhanced sampling molecular dynamics techniques, we have computed the free energy of dimerization in the temperature range $500-1680$~K. Initial structures used in this study were obtained from experimental data. The results show that although the effects of shape, presence of oxygen, mass, and internal bonds are tightly intertwined, and their relative importance changes with temperature. In general, mass and the presence of rotatable bonds are the best indicators. 
The results provide knowledge on the inception step and the role that particle characteristics play during inception. In addition, our study highlights the  fact that current models that use stabilomers as monomers for physical aggregation are overestimating the importance of this process during particle nucleation. 
\end{abstract}


\end{frontmatter}

\ifdefined \wordcount
\clearpage
\fi

\section{Introduction}

Polycyclic aromatic compounds (PAC) can play a critical role during the transition from gas-phase species to particles in combustion. Their presence has been detected using various experiments and techniques, and physical and chemical models have been proposed to describe their growth into nuclei for nanoparticles. (See Review and references therein). 
Besides chemical mechanisms that involve polycyclic aromatic hydrocarbons (PAH) and PAH radical radicals interacting with aryl radicals forming cross-link~\cite{2008_DAnna_ParticleInceptionLaminar,2009_DAnna_Combustionformednanoparticlesa,2001_DAnna_reactionpathwaynanoparticle,1985_Frenklach_Detailedkineticmodeling,2000_Richter_Formationpolycyclicaromatic,2001_Ciajolo_relationultravioletexcitedfluorescence,2003_Allouis_Monitoringfuelconsumption,2004_Violi_KineticMonteCarloa,2014_Lai_Stochasticatomisticsimulation}, a physical model that involves the interactions of PAH of moderate size stacking together via intermolecular forces, such as electrostatic force and dispersion, has also been advanced~\cite{2002_Schuetz_NucleationsootMoleculara,2011_Chung_Pericondensedaromaticsaliphatic,2008_Herdman_IntermolecularPotentialCalculations,1991_Miller_kineticspolynucleararomatica,1991_Frenklach_Detailedmodelingsoot,2012_Totton_quantitativestudyclustering,2013_Elvati_Thermodynamicspolyaromatichydrocarbon}. 
The latter has been the focus of various experimental and computational efforts, but uncertainties still remain in terms of the molecular size required for two PAH molecules to form a stable dimer, the effect of temperature, and chemical structure of the species (alkylated PAHs versus highly condensed PAHs). 
However, still uncertainties remain~\cite{2011_Chung_Pericondensedaromaticsaliphatic,2012_Totton_quantitativestudyclustering,2010_Sabbah_ExploringRolePAHs,2011_Grotheer_Photoionizationmassspectrometry,2015_Johansson_Sootprecursorformation}.
To add to this picture, recent work from our group has highlighted the importance of oxygen chemistry during the growth of PAC, leading to the formation of compounds that contain oxygen. These species were identified in a wide variety of combustion conditions, suggesting generic formation mechanisms~\cite{2017_Elvati_Oxygendrivensoot,2016_Johansson_Formationemissionlarge,2017_Dillstrom_effectreactionmechanisms,2017_Johansson_Radicalradicalreactions}.

The presence of oxygenated hydrocarbons can have a significant influence on particle structures and mechanisms of growth. Indeed, oxygen chemistry can overshadow the hydrogen-abstraction-C2H2-addition (HACA) mechanism~\cite{1991_Frenklach_Detailedmodelingsoot,2002_Frenklach_Reactionmechanismsoot} in some conditions, leading to the formation of species with a high content of oxygen atoms. The presence of oxygen atoms on PAC can also influence the tendency of these compounds to form dimers since oxygen should tend to stabilize the assembly because of dipole-dipole interactions. 
With this knowledge in mind, the aim of this work is to determine the role that oxygenated aromatic compounds play during the physical growth of PAC, with the aim of characterizing their tendency to form dimers as function of the oxygen bond (heterocycles, hydroxyl-PAHs, oxy-PAHs) temperature of the system and mass of compounds.

\section{Methodology}
\label{sec:Methodology}

For all the simulations, we employed Not Another Molecular Dynamics (NAMD) program~\cite{2005_Phillips_Scalablemoleculardynamics} in combination with the PLUMED plugin~\cite{2014_Tribello_PLUMEDNewfeathers} for free energy (FE) calculations. 
Molecules were minimized and equilibrated in a canonical ensemble for 1~ns before starting well-tempered Metadynamics~\cite{2008_Barducci_WellTemperedMetadynamicsSmoothly} simulations.
Temperature was kept constant by employing a Langevin thermostat~\cite{1978_Schneider_Moleculardynamicsstudythreedimensional} with a time constant of 0.1~ps.
The general Chemistry at Harvard Macromolecular Mechanics (CHARMM) force field~\cite{2010_Vanommeslaeghe_CHARMMgeneralforce} was used to model atomic interactions and charges. 
Atom-atom non-bonded interactions were cut at 5~nm after being tapered to zero between 4.8 and 5~nm by using the X-PLOR switching function.
Each FE profile is built as the average of three to five independent 100-ns runs, in which the gaussian-shaped bias (height of 0.1~kcal/mol, width of 0.04 nm) was deposited every 100~fs. 
The bias factor ($bf$) was set so that $bf \cdot k_BT = 130$~kJ/mol.

Overall, including homo- and hetero-dimerization, we reconstructed more than 500 free energy surfaces as function of the molecules center of mass distance (COM).
Based on the shape of the free energy curves, specially the location of the transition states and the dimer local minima, we defined the dimerization state as all the conformations for which the distance between the COM of the molecules was in the range $0.35-0.75$~nm. 
Including shorter distances was not possible due to increase in error of poorly sampled distances, but due to their very high values of the FE the contributions of shorter interval is negligible. 
Moreover, as the transition state (local maximum) can be as close as 0.8~nm, we were precluded from choosing larger intervals. 
To monomer state was instead chosen to include all the states with molecular COM distance above 3.8~nm (up to the free energy right margin at 4.0~nm) as, in all the gaseous systems, the intermolecular interactions were weaker than $k_BT$ already at 3.5~nm.  
The dimerization propensity, is computed as the FE difference ($\Delta\Delta$A) between the dimer-state and the monomer-state.

\begin{figure*}[htb]
\centering
\includegraphics[width=14.4 cm]{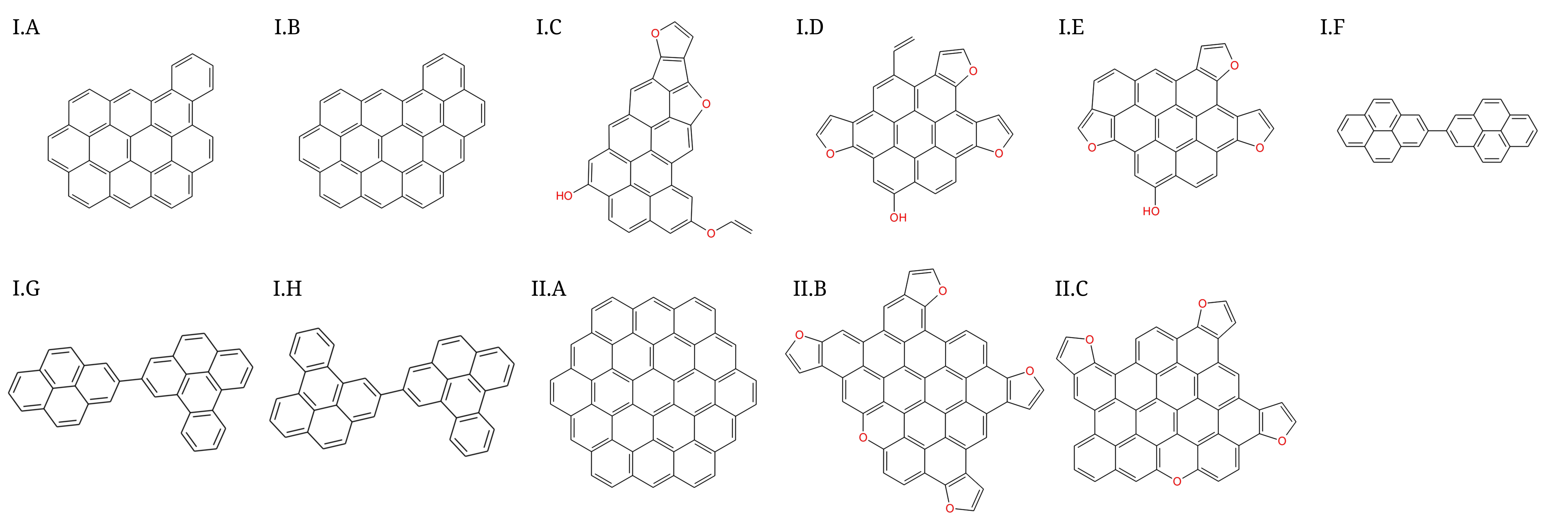}
\caption{Structural formula of the compounds considered in this work.}
\label{fig:structures}
\end{figure*}

\begin{table}[htb]
\caption{Selected properties of the molecules modeled in this work.
Mass is the molecular mass factoring in the natural isotopic abundance; O/C is the ratio between the number of oxygen and carbon atoms, and $R_{gyr}$ is the radius of gyration.}
\centering
\begin{tabular}{llccc}
\toprule
Label & Formula       & Mass (u) & O/C  & $R_{gyr}$ (\AA)\\ 
\midrule
I.A   & \ce{C36H16}   & 448.524 & 0.000 & 4.062 \\
I.B   & \ce{C38H16}   & 472.546 & 0.000 & 4.168 \\
I.C   & \ce{C32H14O4} & 462.460 & 0.125 & 4.584 \\
I.D   & \ce{C32H14O4} & 462.460 & 0.125 & 4.050 \\
I.E   & \ce{C32H12O4} & 460.444 & 0.125 & 4.072 \\
I.F   & \ce{C32H18}   & 402.496 & 0.000 & 5.049 \\
I.G   & \ce{C36H20}   & 452.157 & 0.000 & 5.164 \\
I.H   & \ce{C40H22}   & 502.172 & 0.000 & 5.264 \\
II.A  & \ce{C54H18}   & 666.738 & 0.000 & 4.755 \\
II.B  & \ce{C50H16O5} & 696.673 & 0.100 & 5.146 \\
II.C  & \ce{C48H16O4} & 656.652 & 0.083 & 4.916 \\
\bottomrule
\end{tabular}
\label{tab:properties}
\end{table}

\section{Results}

The structures identified for this study were chosen using experimental data on mass spectra and O/C ratios obtained in a premixed flame of ethylene characterized via synchrotron-coupled VUV aerosol mass spectrometry and X-ray photoelectron spectroscopy (XPS)~\cite{2016_Johansson_Formationemissionlarge}. 
Mass spectra highlighted the presence of oxygenated compounds in flame and XPS identified the types of functional groups, \ie \ce{C-OH}, \ce{C-O-C}, and \ce{C=O}. 
Starting from these experimental evidences, we have identified a pool of aromatic compounds, whose chemical structures are reported in Fig.~\ref{fig:structures}. 
Since dimerization of compounds of small masses is not a likely phenomenon at flame temperatures, we identified our smallest molecule as 402~u. 
Series \textrm{I} includes compounds in the mass range $450-470$~u; Series \textrm{II} is composed of masses between $656-696$~u, close to the mass of circumcoronene. 
The pool of molecules includes large oxygen heterocycles compounds (furans), ethers, aromatics with hydroxyl functionalities, and polycyclic aromatic hydrocarbons. 
In addition to highly condensed rings (I.A, I.B, II.A), we incorporated aromatic aliphatic linked hydrocarbons, constituted by aromatic molecules linked by a $\sigma$-bond (I.F, I.G, and I.H).  
The importance and relevance of these compounds has been reported using experimental and computational techniques~\cite{1998_Ciajolo_Spectroscopiccompositionalsignatures,2007_Chung_Insightsnanoparticleformation,1992_DAlessio_Precursorformationsoot}.

A selection of their properties is listed in Tab.~\ref{tab:properties}. 

In this study we consider three temperatures: 1680~K, which is the temperature in the location of the ethylene premixed flame where bimodal particles were detected~\cite{2017_Elvati_Oxygendrivensoot}, 1000~K, and 500~K, which is below the boiling point of all the molecules considered in our analysis.

\subsection{Homodimerization}

\begin{figure}[!htb]
\centering
\includegraphics[width=6.7 cm]{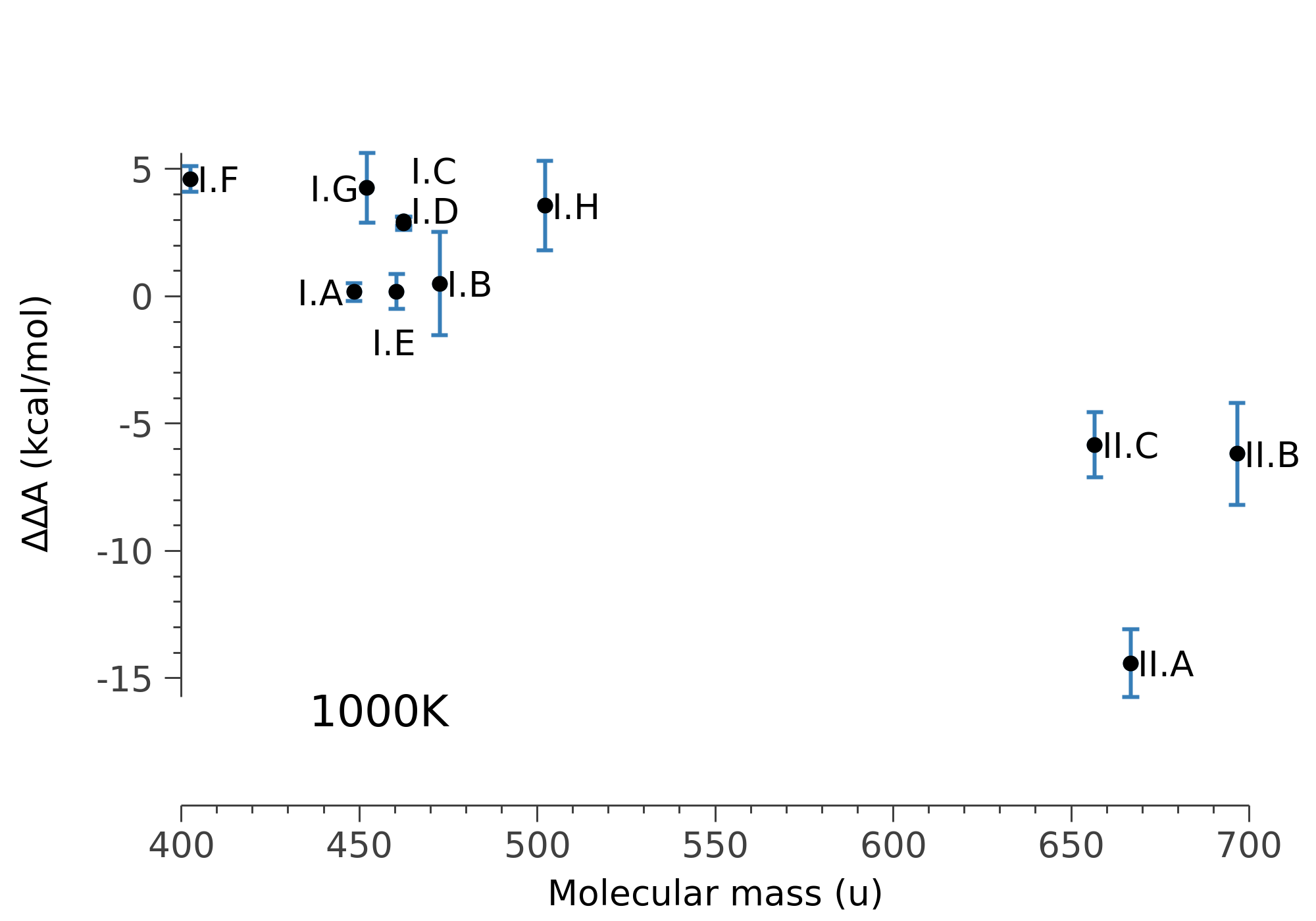}
\caption{Homo-dimerization propensity at 1000~K for the molecules listed in Fig.~\ref{fig:structures}.}
\label{fig:allmass}
\end{figure}

\begin{figure}[!htb]
\centering
\includegraphics[width=6.7 cm]{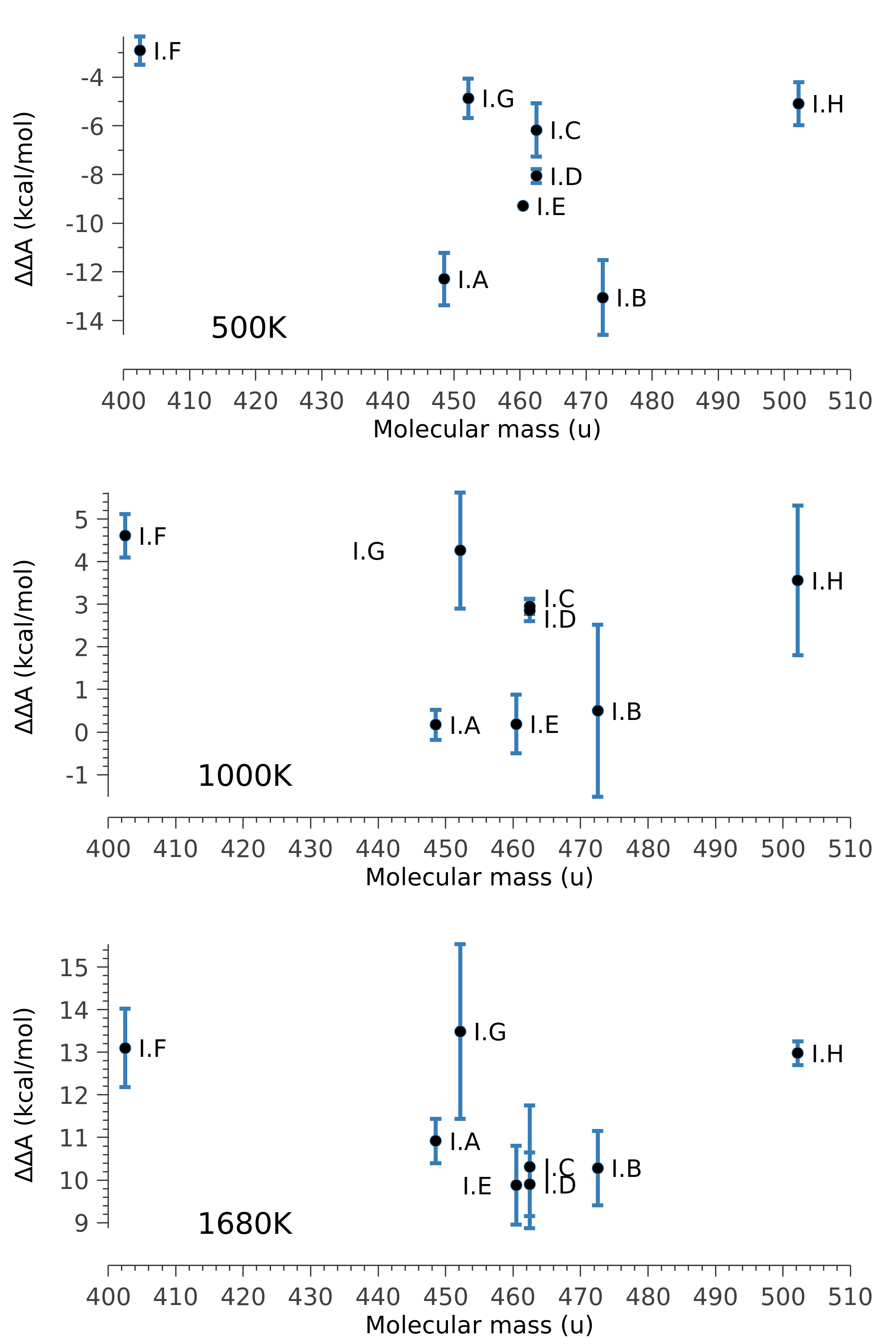}
\caption{Homo-dimerization propensity at different temperatures for molecules the group \textrm{I} as function of their mass.}
\label{fig:Imass}
\end{figure}

Figure~\ref{fig:allmass} shows the dimerization propensity (see Sec.~\ref{sec:Methodology} for the exact definition) for homomolecular systems as function of the mass of the monomers at 1000~K. 
The results show a disorganized clustering in two major groups: one for the molecules in Series~\textrm{I} and one for the molecules in Series~\textrm{II}. The same trends are observed at the other temperatures analyzed. 
Within each group there is, however, only a minimal structure and the lack of a clear monotonic trend demonstrates that mass is only a partially good predictor of the stability of dimers.
Indeed, the dimers of highly fused polycyclic hydrocarbons, \eg I.A, I.B, and II.A, are more stable than other compounds with relatively higher masses, as can be observed in Fig.~\ref{fig:Imass}: for example, I.G (452~u) and I.H (502~u) are less stable than I.A (448~u) over the temperature range studied.
A careful analysis of the data shown above, suggests that there are several phenomena that contribute to these results.

\begin{figure}[!htb]
\centering
\includegraphics[width=6.7 cm]{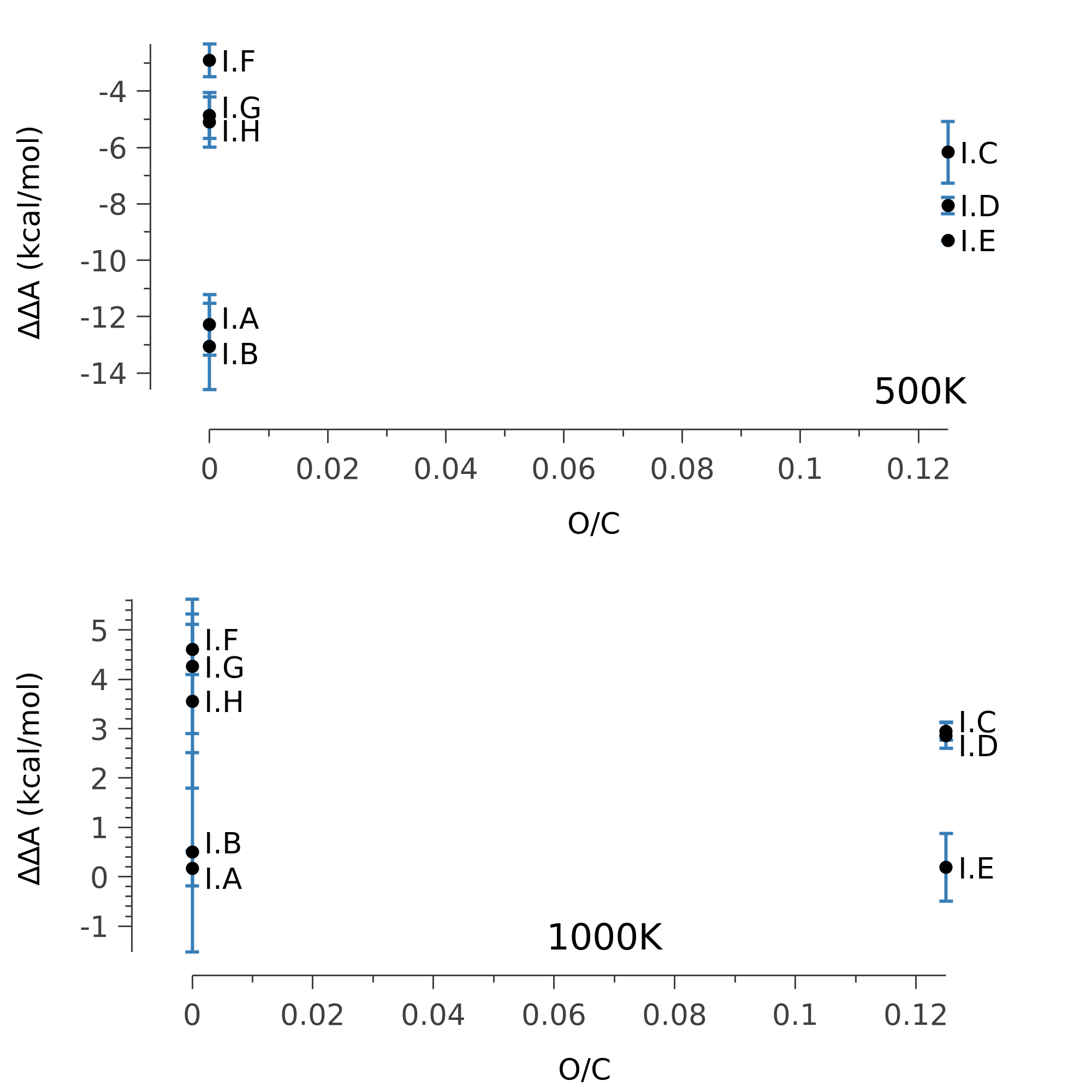}
\caption{Homo-dimerization propensity at 500~K and 1000~K as a function of their oxygen/carbon ratio.}
\label{fig:Iox}
\end{figure}

Figure~\ref{fig:Iox} reports the difference of free energy of dimerization as function of the oxygen content of the molecules at 500~K and 1000~K. 
The data points are broadly clustered in three groups.
The first, composed of the compounds I.C (462~u), I.D (462~u), and I.E (460~u), which have the same mass and O/C ratios (0.125), and are less stable than I.A (448~u) and I.B (472~u), which constitute the second group.
This difference indicates that the presence of oxygen slightly destabilizes the dimers, an effect possibly caused by the increased repulsive electrostatic interactions of oxygenated molecules compared to pure aromatic hydrocarbons.
The third group is formed by I.F (402~u), I.G (452~u), and I.H (502~u), which are less likely to dimerize than the other two groups.
The low dimer stability of the third group can be either due to the presence of a $\sigma$-bond, which introduces an internal rotatable bond (IRB) that interferes with the formation of clusters, or due to the elongated shape of the molecules in this group.
With the goal to clarify the effect of shape, we analyzed the correlation between homo-dimerization propensity and different geometrical descriptors. 
In the end, we found that the radius of gyration ($R_{gyr}$) is able to describe some of the trends we observed above better than other standard descriptors based on principal axis of inertia. One possible reason for this result is that at all the temperatures considered, the molecules are in a highly excited rotational state and $R_{gyr}$ roughly approximates the geometry of the molecules in those conditions.

\begin{figure}[!htb]
\centering
\includegraphics[width=6.7 cm]{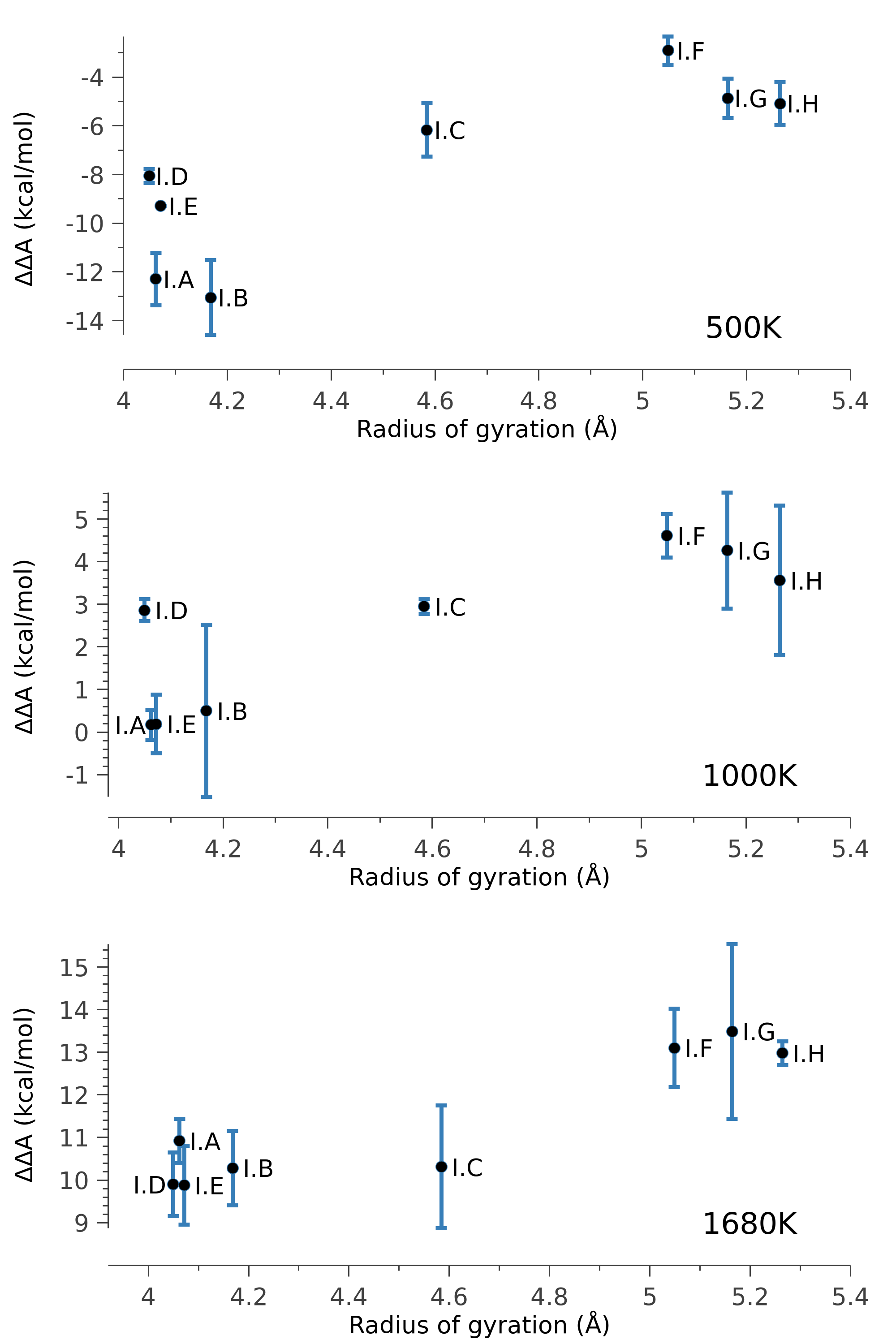}
\caption{Homo-dimerization propensity at different temperatures as a function of the radius of gyration of the monomers.}
\label{fig:Igyr}
\end{figure}

In Fig.~\ref{fig:Igyr} we plotted the dimerization propensities as function of the radius of gyration of the monomers. 
The results show that generally the stability of dimers decreases as the radius of gyration increases.
However, the physical reasons behind this trend are not as evident.
The reason is related to the fact that is difficult to separate the contributions of different competing effects.
To illustrate this issue, we compared I.C, I.D, and I.E: these compounds have the same O/C ratio, and almost identical mass.
However, I.C has a higher $R_{gyr}$ than I.E (and I.D) and simultaneously a lower dimerization propensity, which seemingly points to an influence of the shape.
When comparing I.E(460~u) and I.D(462~u) is clear that the difference in dimer stability can also have another origin, as I.E and I.D do not differ in shape or $R_{gyr}$ but they do in aggregation propensity.
A comparison between the structures of the molecules does not provide a clear answer, except suggesting that the presence of the ethenyl chain does not facilitate the dimerization.

Other phenomena can be inferred by comparing the trends among similar molecules in Fig.~\ref{fig:Igyr}.
For example, within the group formed by I.F (402~u), I.G (452~u), and I.H (502~u), we can identify an internal trend that is related to the mass. 
Clearly, other factors (\ie O/C and IRB) being equal, the stabilization due to the mass dominates over the loss of stability due to shape.
The same effect is also visible when comparing I.A and I.D, which have very close values of $R_{gyr}$ ($\approx$~4) but different masses, although in this case also I.D is further destabilized by the presence of furanyl rings.

\subsection{Heterodimerization}

While the analysis of the homo-molecular interactions are insightful, hetero-molecular interactions are more relevant for real systems.
Therefore, to verify how the trends observed in the previous section are reflected when different molecules aggregate, we simulated the dimerization of all the possible pair combinations of the monomers shown in Fig.~\ref{fig:structures}.

\begin{figure}[!htb]
\centering
\includegraphics[width=6.7 cm]{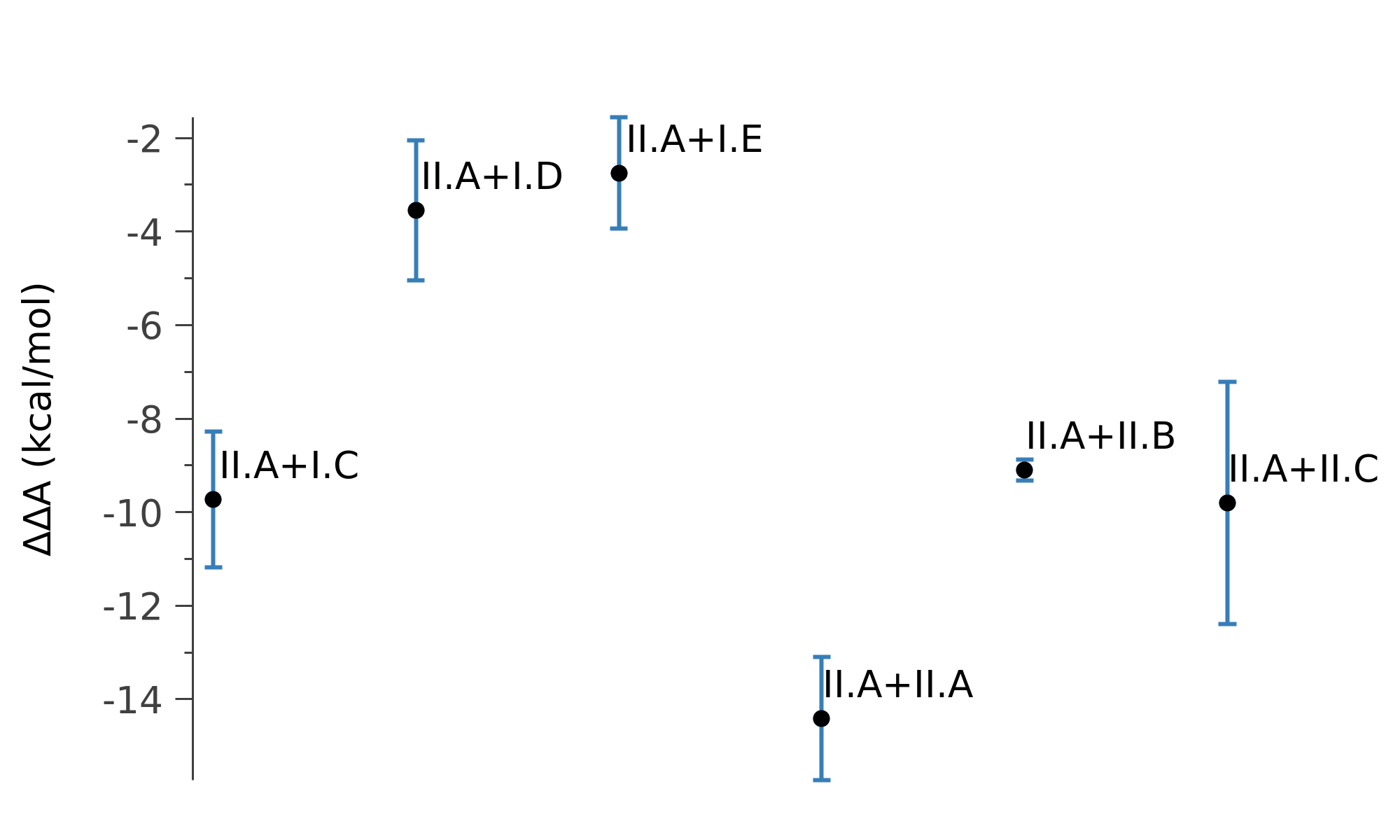}
\caption{Dimerization propensity of different molecules with II.A at 1000~K}
\label{fig:IIAHetero}
\end{figure}

In Fig.~\ref{fig:IIAHetero} we show the results for the dimerization of circumcoronene (II.A, 667~u) with I.C (462~u), I.D (462~u), I.E (460~u), II.B (696~u), and II.C (656~u) as well as the homo-dimerization propensity of II.A.
Although II.A and II.C have similar masses and  $R_gyr$, the dimer II.A-II.C is less stable than the homo-dimer II.A-II.A. 
This difference is likely due to the presence of oxygen, in agreement with the previous results. 
This phenomenon is more evident when we consider the dimer II.A-II.B. 
Although II.B (696~u) is much heavier than II.A (666~u), and therefore we could expect a greater stability as compared to II.A-II.A dimer, the presence of oxygen overshadows the effect of mass resulting in a decrease of the dimerization propensity.
The II.A-II.B dimer is even less stable than II.A-II.C (about -9.1 kcal/mol and -9.8 kcal/mol, respectively) suggesting that the stabilization due to the increase in mass associated with the addition of a single furanyl ring is more than offset by the electrostatic repulsion.
Interestingly, this trend points to the fact that the presence of oxygen is not only a destabilizing factor when both monomers have oxygen atoms, but also when they interact with polycyclic aromatic hydrocarbons. 

The results of the interactions of II.A with smaller molecules show also interesting trends. 
I.C and I.D have the same mass, same oxygen content, same number of 6-membered and 5-membered rings, same number of side chains, but they show a rather relevant difference in stability (-9.73 kcal/mol and -3.55 kcal/mol)
The high dimerization propensity of I.C (compared to all the other molecules in Series \textrm{I}) with larger molecules can be observed even at 1680~K and when interacting with II.B and II.C. This is an opposite result compared with the trends observed for homo-dimerization.
We currently have no explanation for this phenomenon, although it should be noted that I.C is the only molecule in the set that has an aliphatic ether chain, as well as some curvature propensity.


\subsection{General observations}
From the above discussion it is clear that there are several effects that are closely intertwined, and it is difficult to separate them, because shape, IRBs, and mass are not completely de-correlated. However, we can make some general observations.

First, the details of the structure of the species that nucleate cannot be ignored. 
It has been suggested, that the mounting evidence of the presence of five-membered rings, aliphatic side chains, and oxygenated groups in soot precursors, does not change the nucleation mechanism that lead to soot formation.
Our results, both pertaining to the dimerization propensity and the change in free energy barriers between monomers and dimers, which directly relates to the kinetic rates of dimerization, show otherwise. [REFS]
While some clear trends can be observed when a specific group of molecules is selected (\eg stabilomers, $\sigma$ bonded polycyclic aromatic hydrocarbons (PAHs)), differences in shape, oxygen content, and IRBs have large impact on nucleation.
Certainly, under specific assumption, \eg irreversible nucleation, these effects can be ignored, but these approximations lack a general physical explanation.

Second, the effects of shape, presence of oxygen, mass, and IRBs are tightly intertwined, and have different importance as well as diverse temperature dependencies, although, as a general trend, they are all dominated by entropic effects at high ($\approx$~1700~K) temperatures.
While classical molecular dynamics is only semiquantitative, our results suggest that mass and the presence of IRBs are the most important characteristics of molecules, followed by the presence of oxygen and side chains, whereas shape seems to play a minor role. 

Third, highly symmetric stabilomers form the most stable dimers.
As such we hypothesize that they are more likely to constitute the soot core, in agreement with several experimental findings.
Other molecules are instead more prone to nucleate later, which will place them closer to, if not on, the surface of particles, where they are more likely to react during soot aging.

\section{Conclusions}
In this work, we used enhanced sampling Molecular Dynamics techniques to analyze how dimerization n in the flame maybe affected when different factors, like the presence of oxygen and molecular shape, are taken into account.
To this end we have selected a variety of molecules and analyzed their propensity to form dimers at different temperatures from 1680~K to 500~K.
Our analysis, shows that there is a complex interplay between several effects where mass and internal rotatable bonds play a key role in determining the stability of the aggregates.
Also the presence of oxygen, affects the dimer propensity, by reducing the molecular cohesion due to electrostatic repulsion although, it should be noted that the force field employed in this study cannot capture the effect of molecular polarizability.

Overall, all the trends show that any deviation from stabilomers structure, leads to a lower dimerization propensity, which is likely the reason why regular PAHs are most commonly found in the soot core.
This result also suggests that current models that are mostly based around the properties of stabilomers are overestimating the importance of the role of physical aggregation in the soot inception.
This work is the last of series of studies by experimental and computational groups that show how that the importance of physical dimerization, at high temperatures, should be drastically reduced and other mechanisms, like radical recombinations or carbonization reactions, should be instead investigated in order to explain experimental data.

\bibliography{Oxy-PAH_Com.bib}
\bibliographystyle{elsarticle-num-PROCI.bst}

\end{document}